\documentclass[11pt,hyper,letter]{JHEP3}
 \pdfoutput=1
 \usepackage{graphicx,amssymb,amsmath,amsfonts,cite}
 \usepackage[english]{babel}
 \usepackage{amssymb}
 \usepackage{amsfonts}
 \usepackage{amsmath}
 \usepackage{graphicx}
 \usepackage{epsfig}
 \usepackage{cite}
 \newcommand{\bea}{\begin{eqnarray}}
 \newcommand{\eea}{\end{eqnarray}}
 \newcommand{\non}{\nonumber \\}
 \newcommand{\CR}{\non\cr}
 \newcommand{\pa}{\partial}

 \newcommand\be{\begin{equation}}
 \newcommand\ee{\end{equation}}

 \newcommand{\s}{{\sigma}}
 \newcommand{\ts}{{\tilde\sigma}}
 {\small {\Large {\large }}}
 
 \newcommand{\al}{\alpha}

 \vspace{18pt}

 \title   {\Large  On bound-states of the Gross Neveu model with \\
 	massive fundamental  fermions }
 \author
 { Yitzhak Frishman${}^1$\footnotemark[1]  and Jacob Sonnenschein${}^2$\footnotemark[2] \\
 	${}^1$\textit {
 		Department of Particle Physics and Astrophysics\\
 		The Weizmann Institute of Science, Rehovot 76100, Israel} \\
 	\\
 	${}^2$\textit{
 		The Raymond and Beverly Sackler School of Physics and Astronomy,\\
 		Tel Aviv University, Ramat Aviv 69978, Israel}\\

 	\footnotetext[1]{E-mail address: \email{yitzhak.frishman@weizmann.ac.il}}
 	\footnotetext[2]{E-mail address: \email{cobi@post.tau.ac.il}}

 }

 \vspace{18pt}
 \abstract
 {In the search for QFT's that admit boundstates, we reinvestigate the two dimensional Gross-Neveu  model, but with massive fermions. By computing the self-energy for the auxiliary boundstate field and the effective potential, we show that
 there are no bound states around the lowest minimum, but there is a meta-stable  bound state
 around the other minimum, a local one. The latter decays 
 by tunneling. We determine   the dependence of 
 its lifetime  on the fermion mass and coupling constant.}


 \preprint{WIS/03/15-JUL-DPPA,  TAUP-2966/13}
 \newpage

 \newcommand{\kslash}{\not k}

\begin{document}
\section{Introduction}
Bound states are a very important ingredient of  physical systems. From hadrons via atoms and molecules and  all the way to solar systems and galaxies  we encounter in nature systems where a bunch of `` elementary objects" are bound together.
Whereas, for most of such systems the binding mechanism is well understood, for hadrons the process of confining together quarks and gluons is still quite far from being fully understood. Moreover, QFT in general has been developed to handle mainly amplitudes of scattering and much less tools have been constructed to analyze the phenomena of bound states.

A landmark development en route to deciphering confinement in four dimensional QCD  has been 't Hooft's solution of two dimensional QCD in the large number of colors $N$ limit\cite{'tHooft:1974hx}. A Bethe-Salpeter equation for the wavefunction of the quark anti-quark bound state was solved and the corresponding spectrum was determined.
It is easy to realize that the same ``magic" cannot be achieved for  QCD in higher, even not three, dimensions. A simpler toy model in three dimensions is the Chern-Simons  (CS) theory coupled to fundamental matter. Tremendous progress has been made in recent years in understanding  various aspects of these systems. In particular we have shown\cite{Frishman:2013dvg},\cite{Frishman:2014cma}  that
 unlike the two dimensional QCD case, the
three dimensional Chern-Simons theory  coupled to fundamental fermions in the large $N$, large level  $K$ limits where $\lambda=\frac{N}{K}$ is kept fixed, does not admit a bound state spectrum. This was done by proving that the 't Hooft like bound state equations do not have solutions.  On the other hand in \cite{Jain:2014nza} it was shown  that the CS theory coupled to scalars in the fundamental representation with quartic self interaction does admit poles in the  scattering  amplitude   which  correspond  to  particle  anti-particle  bound
states in the singlet channel.  Furthermore,  it was shown \cite{Aharony:2012ns} that this theory in the so called ``Wilson-Fisher" limit is equivalent to the theory of fundamental fermions coupled to a CS gauge theory.   This equivalence holds for  a range of parameters that does not include  the region where the S matrix has poles so there is no contradiction between the no bound states of \cite{Frishman:2014cma} and the poles of \cite{Jain:2014nza}.\footnote{The possibility of having a massless bound state was discussed in  \cite{Bardeen:2014qua}}
These findings raise a natural question of whether the theories  of fundamental matter with quartic interaction both for bosons and fermions but with no coupling to the CS gauge fields admit bound states. This question, in various dimensions, is the subject of the current paper and a following one\cite{FS}.

Unlike the method we have used in   \cite{Frishman:2013dvg} and \cite{Frishman:2014cma}, in this paper we use a much simpler method. Following the gross Neveu model [GN] \cite{Gross:1974jv}, we formulate the quartic interactions by invoking a singlet auxiliary field which trivially  upon integrating it out yields the ordinary formulation. We compute (exactly in large $N$) the self-energy of the auxiliary field, but this time with non-zero fermion mass, and check for what domain of the parameters of the theory it admits poles.  
It turns out that the effective potential has two minima. There are no bound states 
around the lower one, while there is one bound state around the other one. This state is  unstable. It decays  by tunneling to the lower minimum, which is the ground state. Its life time dependence on the mass and the coupling constant is determined.
In the limit of infinite coupling this state becomes massless and the binding energy is twice the fermion physical mass. It is worth mentioning that the binding at threshold for the massless case is the leading result in large N, but  at finite N  it was found out in \cite{Dashen:1975xh}  that $\frac{M_\sigma}{2M_F}= \cos\left[\frac{\pi}{N-2} \right]$ so that  there is a binding energy at order $O(\frac{1}{N^2})$ (See also \cite{Moshe:2003xn}
eq 4.27).

The GN model and its  generalizations  have been  thoroughly investigated  from various different points of view. For review papers and references therein see  \cite{Rosenstein:1990nm} in 2+1 dimensions and \cite{Moshe:2003xn}. The massive GN model has attracted less attention than the massless one, however certain properties of the massive were also examined for instance in \cite{Dashen:1975xh} \cite{Feinberg:1996kr}, where the binding energies for all the antisymmetric tensor multiplets were computed.

The paper is organized as follows. In section $\S 2$ we describe three methods of handling bound states in QFT, and in particular we elaborate on the auxiliary field method.  In the next section we review the self energy of the auxiliary field.  In section $\S4 $
we calculate the effective potential, for the case of a massive fermion.
 We show that for a certain range of the  coupling constant the system admits one
 quasi-stable bound state. In the last section we summarize the results of  the various different  cases and present some open questions.

\section{Probing bound states}
	Quantum field theories are equipped with efficient tools to study scattering processes, but the toolkit for analyzing bound states is much more limited. Among the methods to address the issue of bound states are:

  1. The identification of poles of the S matrix,

  2. Solving the bound state wave-function,  like 't Hooft in two dimensions, and

  3. Studying the propagator of an auxiliary field, that upon the use of the equations of   \indent
  motion equals to a bound state of two underlying fields.

\noindent
 Let us now elaborate on the latter.
Four Fermi interactions  like that of the Gross-Neveu model is known to be renormalizable in two space-time dimensions. However in the large $N$ limit such an interaction is renormalizable also in higher dimensions\cite{Rosenstein:1990nm}. For the class of models with quartic interactions of fields, either  fermionic or bosonic, one can  examine the question of whether the model admits a bound state by examining the propagator of an auxilary field. The latter is introduced in the following form
\be
{\cal L}( \Phi,\pa_\mu\Phi,  \ts) = {\cal L}_0(\Phi,\pa_\mu \Phi) + \frac12\mu^2 \ts^2 - g \ts \Phi^\dagger\Phi
\ee
where $\Phi$ is a field, either bosoic or fermionic, in the fundamental representation of $SU(N)$.
The equation of motion of $\ts$ reads
\be\label{EOMts}
\ts = \frac{g}{\mu^2}\Phi^\dagger\Phi
\ee
We take now the following limit
\be
g^2\rightarrow \infty \qquad \mu^2\rightarrow \infty \qquad  \frac{g^2}{\mu^2}\equiv \tilde g^2 \ \ fixed
\ee
In fact taking the limit of $\mu^2\rightarrow \infty$ is needed if one adds a kinetic term for  $\ts$.
Using the equation of motion of $\ts$ we obviously get  a quartic interaction in the action

\be
{\cal L}( \Phi,\pa_\mu\Phi)= {\cal L}_0(\Phi,\pa_\mu \Phi) - \frac12 \tilde g^2 (\Phi^\dagger \Phi)^2
\ee
Note that this emerged from a filed $\ts$ that is actually a proper  limit of an infinite mass tachyon field. This is important for the scalar case, to ensure positivity of the energy.

The tree level  and full propagators   of the auxilary field $\ts$ are
\be\label{tach sigma}
D_\ts^{0}(k) =i  \qquad D_\ts(k) =\frac{i}{1- i\Pi(k)}
\ee
where $\Pi(k)$ is the self energy of the auxiliary field.
From the outcome of the equation of motion (\ref{EOMts})  it follows that $\ts$ is the field that describes a boundstate of $\Phi^\dagger\Phi$. Thus, a boundstate exists if

 {\bf  $D_\ts(k)$ has a pole in the domain of $ 4M_\Phi^2 > k^2 > 0 $}.

where $M_\Phi$ is the physical mass of the particle created by the field $\Phi$.
\noindent If on the other hand

{\bf There is a pole in the  domain $(-k^2)>0 $, then the theory suffers from a Landau pole.}

\section{The self energy of $\sigma$}
We start   with the  Lagrangian density
	\be\label{3dF}
	{\cal L} = \bar{\psi} ( i\gamma^\mu\pa_\mu -m)\psi - \frac{1}{2}\s^2 - g\s \bar{\psi\psi}
	\ee
	where $\psi$ is 2d Dirac fermion in the fundamental representation of $U(N)$ and $\s$ is a scalar which is a $U(N)$ singlet.
    The $\gamma$ matrices in 2d can be expressed in terms Pauli matrices $\gamma_0=\sigma_1,\gamma_1=i\sigma_2  $. Note that this model differs from the orginal GN model  by the fact that  here the fermions are  massive.
	Since $\s$ does not have a dynamical term its free propagator is a constant
	\be
	D_\s^0 (k) = -i
	\ee
	The fermion {\bf free} propagator is given by
	\be
	S_\psi^0(k)=\frac{i}{\kslash -m}
	\ee
	Next we proceed to compute the scalar self energy.  We do it in the large $N$ limit for which
	\be
	N\rightarrow \infty \qquad g^2 N \equiv \lambda\  fixed
	\ee
	where $\lambda$ is dimensionless.
	The full propagator of $\s$ is given by 
	\be\label{propagator}
     D_\s (k) = \frac{-i}{1+i\Pi(k)}
	\ee
	where

\bea\label{pik}
\Pi(k)&=& \lambda\int \frac{d^2 q}{(2\pi)^2} Tr\left [ \frac{i}{[\gamma^\mu q_\mu -m]}   \frac{i}{[\gamma^\nu (q-k)_\nu -m]} \right ]\CR
&=& - \lambda\int \frac{d^2 q}{(2\pi)^2}  \frac{2(q^2-k\cdot q + m^2)}{(q^2-m^2)[(k-q)^2-m^2]} \CR
\eea
Note that in the large $N$ limit, the full result is given by the {\bf one loop} contribution.

The full propagator in \ref{propagator} is different than in \ref{tach sigma}, since in the latter 
we had a tachyonic $\sigma$.

As is common we introduce a Feynmann variable $\alpha$ so that
\bea
    \Pi(k)&=&-2 \lambda\int_0^1 d\al\int \frac{d^2 q}{(2\pi)^2}  \frac{q^2-k\cdot q + m^2}{[q^2 +\al k^2-m^2-2\al k\cdot q]^2}\CR
	      &=& -2 \lambda\int_0^1 d\al\int \frac{d^2 q}{(2\pi)^2}  \frac
	           {(q+\al k)^2- (\al k+ q)\cdot k  + m^2}{[q^2 +\al (1-\al)k^2-m^2]^2}\CR
	      &=& -2 \lambda\int_0^1 d\al\int \frac{d^2 q}{(2\pi)^2}  \frac{q^2 + m^2- \al(1-\al) k^2   }{[q^2+\al(1-\al)k^2 -m^2]^2}\CR
	      &=& 2i \lambda\int_0^1 d\al\int \frac{d^2 q_E}{(2\pi)^2}  \frac{q_E^2-[m^2 -\al(1-\al) k^2]}{[q_E^2 +m^2-\al(1-\al)k^2]^2}\CR
\eea

Incorporating now a dimensional regularization with $D=2-2\epsilon$ we get
\bea
\Pi(k^2) &=&\frac{i\lambda}{2\pi}\int_0^1 d\al\left[\frac{1}{m^2}[m^2-\al(1-\al)k^2]\right]^{-\epsilon}\int_0^\infty \xi^{1-2\epsilon} d\xi \frac{\xi^2-1}{(\xi^2+1)^2}\CR
 &=&\frac{i\lambda}{2\pi}\int_0^1 d\al\left[\frac{1}{m^2}[m^2-\al(1-\al)k^2]\right]^{-\epsilon}\left[\frac{1}{\epsilon} -1+O(\epsilon) \right ]\CR
 &=& \frac{i\lambda}{2\pi}\left[ \frac{1}{\epsilon} -1\right]- \frac{i\lambda}{2\pi}\int_0^1 d\al \ln \left[\frac{1}{m^2}[m^2-\al(1-\al)k^2]\right]\CR
\eea
Define now the renormalized coupling
\be
G_R^2=\frac{\lambda}{1-\frac{\lambda}{2\pi}\left(\frac{1}{\epsilon} -1) \right)}
\ee
A pole in the propagator appears when
\be\label{pole}
1 + \frac{G_R^2}{2\pi}\int_0^1 d\al \ln \left[\frac{1}{m^2}[m^2-\al(1-\al)k^2]\right] =0
\ee
In fact as will be shown below the condition for the pole is determined by this equation when we replace
$$m\rightarrow M_F= m+ g\sigma_m$$
where $\sigma_m$ is the minimum of the effective potential.

The requirement of  no Landau pole is obeyed for
\be
\frac{G_R^2}{\pi}\geq 1
\ee
The system admits one bound state.
Note that in this case the self-energy vanishes at $k^2=0$, so no subtraction needed.

Of course to get the full effects, we need to compute the $\sigma$ potential,
as was done by Gross-Neveu \cite{Gross:1974jv}, but with a finite mass $m$ this time.

 We have used dimensional regularization here, as will be more convenient in higher dimensions. However in the next section we will use the subtraction method, like in
the GN paper.

\section{The effective potential}
Due to translational  invariance the vacuum expectation value of the field $\sigma(x)$ has to be $x^\mu$ independent, namely a constant. We denote it by $\s_c= <0|\s|0>$. The generating functional is given by
\be
\Gamma(\s_c)= \int d^Dx V(\s_c)
\ee	
where $	V(\s_c)$ is the effective potential.
Since $\Gamma$ is the generating functional of the  one particle irreducible (1PI)  $n$ point function the effective potential is thus
\be
V(\s_c)= \sum_n\frac{1}{n!}(\s_c)^n \Gamma_n(0,0,...,0)
\ee
where $\Gamma_n(0,0,...,0)$ is the sum of all the 1PI Green's function with $n$    external lines of $\s$
at zero momentum.
At tree level it is obvious that $V(\s_c= \frac12 (\s_c)^2$.
To leading order in $\frac{1}{N}$  we have to sum all the one loop graphs with a UV cutoff $\Lambda$. Therefore the effective potential is
\be
V(\s_c)= \frac12 (\s_c)^2 -iN \sum_{n=1}^{\infty}\frac{( g\s_c)^n}{n}\int\frac{d^Dk}{(2\pi)^D} Tr\left(\frac{1}{\kslash-m} \right )^n
\ee
Performing the infinite sum we get
\be
V(\s_c)= \frac12 (\s_c)^2 +iN \int\frac{d^Dk}{(2\pi)^D} Tr\left( \frac{1}{2}ln(1-\frac{g\s_c}{\kslash-m})^2 + \frac{g\s_c}{\kslash-m} \right )
\ee
We can now make a change of variables $k^\mu\rightarrow -k^\mu$. We then take a half the sum of the original and modified integrals to get
\be
V(\s_c)= \frac12 (\s_c)^2 +\frac{iN}{2} \int\frac{d^Dk}{(2\pi)^D}Tr \left( \frac{1}{2}ln(1-\frac{(g\s_c+m)^2- m^2}{k^2-m^2})^2 + \frac{2 m g\s_c}{k^2-m^2} \right )
\ee
The Tr would give a factor of $2^{[\frac{D}{2}]}$, where $[\frac{D}{2}]$ is the
integer part of $\frac{D}{2}$ . Going to Euclidean space, and integrating over the angles of the D dimensional sphere we get
\be
V(\s_c)= \frac12 (\s_c)^2 -K(D) \int_0^\infty x^{D-1} dx  \left( ln(1+\frac{(g\s_c+m)^2- m^2}{x^2+m^2}) - \frac{2 m g\s_c}{x^2+m^2} \right )
\ee
where
$K(D)= \frac{N}{C (2\pi)^{(\frac{D}{2})}{\Gamma(\frac{D}{2})}} $, with $C=1$ for even D
and $C=\sqrt 2$ for odd D. The area of the unit sphere in D dimensions, namely
$\frac{2 \pi^{(\frac{D}{2})}}{\Gamma(\frac{D}{2})}$, was used.

We can  evaluate the integral in general $D$ dimensions. The explicit expression is lengthy, and will be given elsewhere.

For D=2 and after introducing a cutoff $\Lambda$ the effective potential reads
\be
V(\s_c)= \frac{1}{2} (\s_c)^2 -\frac{ N }{4\pi}\left[(g\s_c )^2
	ln\frac{\Lambda^2}{m^2} - (g\s_c+m)^2  ln\frac{(g\s_c+m)^2}{m^2}+ (g\s_c+m)^2- m^2 \right]\ee
Due to the divergence, renormalize by subtracting 
$$V(\s_c)\rightarrow V(\s_c)-\frac{1}{2} C_R (\s_c)^2$$
Following \cite{Gross:1974jv} we now that
\be\label{renormalization}
\frac{\pa^2 V}{\pa {\s_c}^2}|_{\s_c=\s_0} =1
\ee
This leads to the relation
$$C_R=-\frac{\lambda}{2\pi}[ln\frac{\Lambda^2}{m^2}-ln\frac{(g\s_0+m)^2}{m^2} -2] $$
From which one gets, for the subtracted potential
\be\label{effpot}
V(\s_c)= \frac{1}{2} \s_c^2 +\frac{ N }{4\pi}\left[(g\s_c )^2 ln\frac{(g\s_c+m)^2}{(g\s_0+m)^2} + (m^2 + 2 mg\s_c) ln\frac{(g\s_c+m)^2}{m^2} -3(g\s_c )^2  -2mg\s_c\right]
\ee
The derivative of the potential is
\be
V'(\s_c)=\s_c +\frac {Ng}{2\pi}\left[g\s_c ln\frac{(g\s_c+m)^2}{(g\s_0+m)^2}+mln\frac{(g\s_c+m)^2}{m^2}-2g\s_c\right]
\ee

We want to find the minimum of the potential.
For the case of $m=0$ the result  agrees  with  \cite{Gross:1974jv}, of course.
For that case the minimum of the potential is at
\be
\s_m=\s_0 e^{(1-\frac{\pi}{\lambda})}
\ee
and the mass of the fermion is $m_F= g\s_m$.
For $m\neq 0$ there is no analytic solution for the minimum $\s_m$. However one can find its approximated value numerically. For instance for the case that $\epsilon=\frac{m}{g\s_0}<<1$ we get that $\frac{\s_m}{\s_0}$ will have, besides $ e^{(1-\frac{\pi}{\lambda})}$, extra
terms of $\epsilon ln\epsilon$ and $\epsilon$, which can be computed directly from
$V'(\s_m)=0$.

\begin{figure}[h!]
		\begin{center}
			\vspace{3ex}
			\includegraphics[width= 80mm]{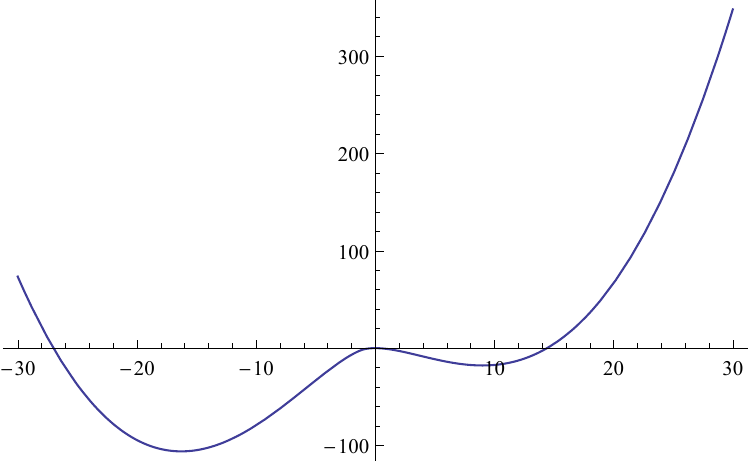}
		\end{center}
		\caption{ $\frac{V g^2}{m^2}$ as a function of $\frac{g\s_c}{m}$ for   $\rho=0.1$   }
		\label{1pIrho}
	\end{figure}

We can rewrite (\ref{effpot}) as
\be
V(\s_c)= \frac{1}{2} \s_c^2 +\frac{ N }{4\pi}\left[(g\s_c + m )^2 ln\frac{(g\s_c+m)^2}{m^2} + 2(g\s_c)^2ln\rho -3(g\s_c )^2  -2mg\s_c\right]
\ee
where $\rho=\frac{m}{g\s_0+m}$. We now choose the scheme where $V'(\s_0)=0$, in which
case the coupling can be eliminated as 
$$\frac{\lambda}{\pi}=\frac{(1-\rho)}{[(1-\rho)+ \rho ln\rho]}$$
Defining $x\equiv\frac{g\s_c}{m}$ and  $y\equiv \frac{V g^2}{m^2}$ the potential now takes the form 
\be
y= \frac12 x^2 + \frac{(1-\rho)}{4[(1-\rho)+ \rho ln\rho]}\left [(x+1)^2 ln( (x+1)^2) + 2x^2 ln\rho -3x^2 -2x\right ]
\ee
In figure \ref{1pIrho} we draw $y$ as a function of $x$.
The first and second derivatives now read
\be
y'= x +  \frac{(1-\rho)}{[(1-\rho)+ \rho ln\rho]} \left [ (x+1)ln( |x+1|) + xln\rho -x\right ]
\ee
\be
y"= 1+ \frac{(1-\rho)}{[(1-\rho)+ \rho ln\rho]} \left [ln( |x+1|) +  ln\rho \right ] 
\ee
We can now  check that the point $\s_0$ where the renormalization condition (\ref{renormalization})  $y''=1$ is obeyed occurs  at $x=\frac{1}{\rho}-1$. At this point $y'=0$, as chosen. This is one of the two local minima of the potential (see figure \ref{1pIrho}). It fact this is the ``false"' vacuum  point that occurs for a positive value of $x$. At $x=0$  there is another extremum point since $y'(x=0)=0$ but that is a maximum point. Since $y \rightarrow \infty$ when $x\rightarrow -\infty$ it implies that there is another minimum for negavtive value of $x$. As can be seen in figure \ref{1pIrho} this is the ``true minimum". This minimum obviously determines the fermion mass 
\be
M_F= g\s_m + m = m( x_m+1)
\ee
It is easy to see that $M_F<0$ since for $x=-1$ $y'>0$.
At the true minimum the second derivative of the potential is 
\bea\label{truemin}
V"(\s_m)&=& 1+ \frac{\lambda}{2\pi}  ln\left[\frac{(g\s_m + m)^2}{(g\s_0+m)^2 }\right ]\CR
 &=& \frac{\lambda}{\pi} \left[ 1-\frac{m}{M_F-m}\ln(|\frac{M_F}{m}|)\right ]\CR
\eea
Where we also used the fact that the first derivative vanishes at $x_m$.

\section{Bound states}
To examine the question of whether the system admits bound-states 
we now go back to (\ref{pik}) the equation for $\Pi(k^2)$. First we renomailze the logarithmic  divergent expression by
 by subtraction at $k^2=0$, namely we get $\hat{\Pi}(k^2)\equiv\Pi(k^2)-\Pi(k^2=0)$

\be
\hat{\Pi}(k^2)=\frac{-i\lambda}{2\pi}\int_0^1 d\al ln\left[1-\al(1-\al)\frac{k^2}{m^2}\right]
\ee
A bound state exists if there is a pole in the $\s$ propagator. 
The condition for the latter is    
\be
\left[\frac{\pa^2 V}{\pa {\s_c}^2}|_{\s_c=\s_m}\right] + \frac{\lambda}{2\pi}\int_0^1 d\al \ln \left[1-\al(1-\al)\frac{k^2}{M_F^2}\right] =0
\ee
Following the structure of the minima of the effective potential we now discuss separately the false vacuum denoted by  $\s_{fm}$ and the true 
 vacuum $\s_{tm}$.
 \subsection{Meta-stable bound state}
  For the false vacuum  we showed that $\frac{\pa^2 V}{\pa {\s_c}^2}|_{\s_c=\s_{fm}}=1$ and hence the condition for having a ``meta stable bound state" is 
 \be
 1 + \frac{\lambda}{2\pi}\int_0^1 d\al \ln \left[1-\al(1-\al)\frac{k^2}{M_F^2}\right] =0
 \ee

Performing the integral the condition takes the form
\be\label{polcond}
 \sqrt{\frac{4M_F^2}{k^2}-1}\  \arctan\left[\frac{1}{\sqrt{\frac{4M_F^2}{k^2}-1}}\right]= 1- \frac{\pi}{\lambda} 
 \ee


We use now the physical mass $M_F$, as obtained from the effective potential
calculated above. We also show there, that $\lambda$ is determined by the
ratio $\rho=\frac{m}{M_F}$, and that for $1>\rho>0$ we have $\frac{\lambda}{\pi}
>1$.
Thus in this region there is one bound state.
It is close to the threshold for small $\rho$, and gets closer to zero mass when
$\rho$ gets close to one from below.

We can also define the following dimensionless measure of the binding energy
\be
\eta \equiv \frac{ 4 M_F^2-k^2}{4M_F^2}
\ee
From the condition (\ref{polcond}) we find that binding at threshold $\eta=0$ occurs for $\frac{\lambda}{\pi}=1$ and the maximal binding $\eta=1$ for
$\frac{\lambda}{\pi}\rightarrow \infty$.

\begin{figure}[h!]
	\begin{center}
		\vspace{3ex}
		\includegraphics[width= 80mm]{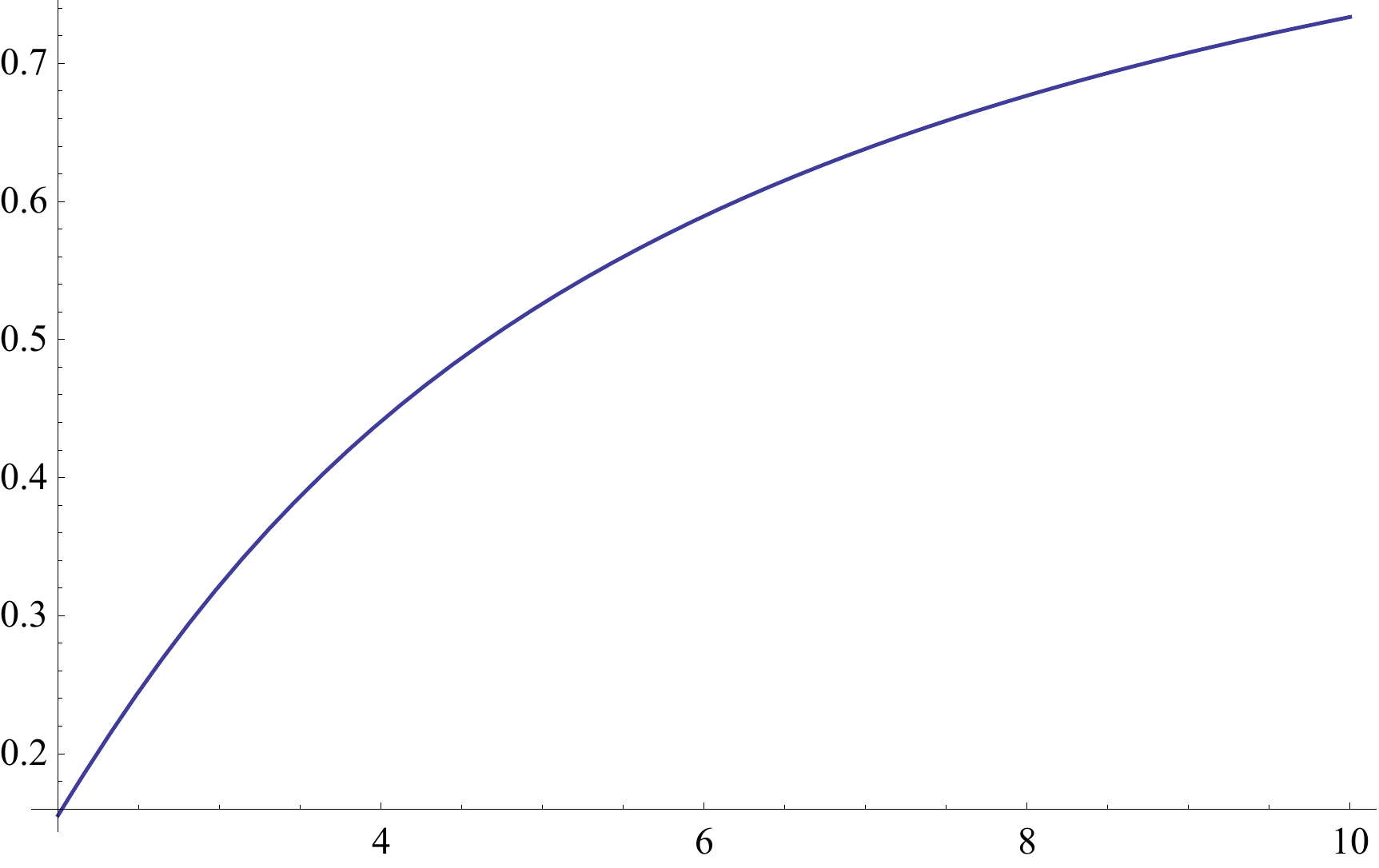}
	\end{center}
	\caption{The  binding energy parameter $\eta$ as a function of the coupling $\lambda$}
		\label{bindingenergy}
	\end{figure}

In figure \ref{bindingenergy} we plot the binding energy parameter $\eta$ as a function of $\frac{\lambda}{\pi}$.

Next we would like to determine the decay width of the meta-stabele state, We follow here the approach of \cite{Coleman:1977py}\footnote{ The fact that this paper was withdrawn by the author is not relevant for our purposes  since the problem was with the   assumption that 
the bounce of minimum action is spherically symmetric, and we deal with a two dimentional case},\cite{Callan:1977pt} about the decay of the false vacuum (see also discussion below). 


The decay width per unit length  for this two dimensional field theory takes the form
\be\label{decaywidth}
\frac{\Gamma}{L} =\frac {B}{h} e^{-\frac{ 2\pi B}{h}}\left [ \frac{det'[-\pa^2+  V''(\sigma)]}{det[-\pa^2+ V''(\sigma_m)]}\right ]^{-1/2}\times \left [ 1+ O(h)\right ]
\ee
where $det'$ is the determinant without the zero modes, and the constant B  which is in fact the Euclidean  action associated with the bounce, is to be 
computed.

 This formula is based on Eq.(3.6) in ref \cite{Callan:1977pt}, 
adopted to the two dimensional case.
This means that the space volume element is now L, and the factor that appears in 
front is $\frac{B}{h}$,   as it is the square root of the factor in the four dimensional case.
 
To compute the decay, we need some approximations.

Following \cite{Coleman:1977py}, we treat the perturbative case of $\frac{m}{g\s_0}<<1$.
So we replace $V$ by $\tilde V(\sigma)$, the ``unperturbed potential",  namely the potential (\ref{effpot}) for the massless case $m=0$ shifted to be zero at the minimum.
 
 We now follow Eq.(4.9) in Ref \cite{Coleman:1977py}, adopted to the two dimensional case.

The suppression factor $B$  which for the 4d case was found to be\cite{Coleman:1977py} $ B=\frac{27\pi^2 S_1^4}{ 2\epsilon^3}$  is now  given by 
\be
B= \frac14 \frac{S_1^2}{\epsilon}
\ee
 $S_1$ is  
\bea\label{s1}
S_1&=& \frac{1}{\s_0}\int_{-\s_{m}}^{\s_{m}} d\s \sqrt{2 \tilde V(\s)} \CR
 &=& {1 \over \sqrt 2} \s_0 \int_{0}^{1} dz \sqrt{z\  ln(z)+1-z}\approx 0.30\s_0 \CR
 \eea
 where we chose the minimum to occur at $\sigma_0$,
 and
the parameter $\epsilon$, that ``measures" the deviation due to the introduction of the fermionic mass, is defined  by approximating the potential as
\be
V(\s,m)\approx V(\s, m=0) + \frac{\epsilon}{2\s_m}(\s - \s_m )
\ee
Using the expression of the potential  (\ref{effpot}) we find that 
\be
\epsilon=  2 m \frac{\s_0}{g} ln\left( \frac {g\s_0}{m}\right )
\ee
Substitute this and (\ref{s1})  into (\ref{decaywidth}) to get 
\be
B\approx 0.011 \frac{\left(\frac{g\s_0}{m}\right)}{ \  ln\left( \frac {g\s_0}{m}\right )}
\ee

So as $m \rightarrow 0$, B tends to $\infty$, thus making the meta-stable state more and 
more long lived.
The other minimum, to leading order in $mln(\frac{g\sigma_0}{m})$, is at

\be
g\s_{tm}=-\left[g\sigma_0 +2mln(\frac{g\sigma_0}{m})\right]
\ee

Now, as in our case the field $\sigma$ has no kinetic term, the expression for the 
determinant in (\ref{decaywidth}) should be modified. 
We will not elaborate on that here, as in any case this will modify the overall coefficient,
but not change the fact that B tends to $\infty$ as $m \rightarrow 0$.

Now to the case of very large $m$ compared to $g\sigma_0$.
In this case the potential at false vacuum tends to a finite depth, 
\be
g^2 V(\sigma_0)\approx -\frac{1}{6}(g\sigma_0)^2
\ee
The true vacuum moves to more negative values,
$$g \s_{tm}\approx -4.6 m$$
and the potential there gets more deep,
\be
g^2 V(\s_{tm})\approx -10.53 m^2 (\frac{m}{g\s_0})
\ee

 The decay of the meta-stable bound state will be slow also in this case, as 
the depth of the false vacuum tends to a constant, while the width of the barrier
grows.
Alas, we do not compute it here.

\subsection{ No stable bound states}
Next we discuss the condition of having a bound state for the system at the true vacuum $\s_{tm}<0$. 
For this case we use the expression found in (\ref{truemin}) for the second derivative of the potential and get that the condition for having a pole reads 
\be\label{nobound}
\frac{\lambda}{\pi} \left[ 1-\frac{m}{M_F-m}\ln(|\frac{M_F}{m}|)\right ] + \frac{\lambda}{2\pi}\int_0^1 d\al \ln \left[1-\al(1-\al)\frac{k^2}{M_F^2}\right] =0
\ee	
Thus we see that the dependence on $\lambda$ cancels out and the condition is 
\be
\left[ 1-\frac{m}{M_F-m}\ln(|\frac{M_F}{m}|)\right ] = \sqrt{\frac{4M_F^2}{k^2}-1}\  \arctan\left[\frac{1}{\sqrt{\frac{4M_F^2}{k^2}-1}}\right]
\ee
One can show that the first derivative is positive also for x=-2, thus getting that
$$|M_F|>m$$
 Now, cancelling the factor $\frac{\lambda}{\pi}$ in (\ref{nobound}), the second term is negative and between 0 and -1, while the first is positive and larger than 1.
Since this condition cannot be fulfilled, there is no stable bound state to the mass deformed GN model
	\section{Summary}
	In this note we address the issue of  bound states in the two dimensional Gross Neveu model with massive fermions. In the original paper it was shown that in the large N limit  for the massless case there is no binding energy, namely, it is a binding at threshold. For the massive case the situation is different, as there is, besides the
	lowest minimum, another one, from which states may tunnel, their life time depending
	of course on the parameters of the system.  
	
    It is shown that there are no bound states above the lowest minimum, while above 
    the other there is one bound state.
    When the coupling constant is $\lambda=\pi$ the binding energy vanishes and when the coupling goes to infinity the binding energy is maximal, namely $2M_F$, twice the fermion physical mass, and the bound state is massless.

We would like to mention that in \cite{Thies:1993jr} \cite{Pausch:1991ee}, based on  the use of a light-cone Tamm-Dankoff approximation, it was found that there is no bound state in the  massive G.N. model\footnote{We would like to thank M. Thies for  bringing this result to our attention and for discussions about it.}.  

	This result is part of a larger investigation of quantum field theories in various dimensions \cite{FS}. Naturally one would like to investigate in a similar manner the Gross Neveu model in three and four dimensions and also the theories with a scalar in the fundamental representation  of the  $SU(N)$ at large $N$.

{\bf Acknowledgements}
We would also like to thank Joshua Feinberg, M. Thies and especially  Moshe Moshe for a very useful correspondence. 
We would  also like to thank   Ofer Aharony  for comments on the manuscript.  
The work of J.S was partially supported by   a center of excellence supported by the Israel Science Foundation (grant number 1989/14),
 by the US-Israel bi-national fund (BSF) grant number 2012383, the Germany Israel
bi-national fund GIF grant number I-244-303.7-2013   and by the ``Einstein Center of Theoretical Physics " at  the Weizmann Institute.


\end{document}